\documentstyle[12pt]{article}

\newcommand{\be}{\small\begin{equation}}
\newcommand{\ee}{\end{equation}\normalsize\vspace*{-0.1ex}}
 \newcommand{\bea}{\small\begin{eqnarray}} 
\newcommand{\eea}{\end{eqnarray}\normalsize\vspace*{-0.1ex}}
 \newcommand{\bdm}{\small\begin{displaymath}} \newcommand{\edm}
{\end{displaymath}\normalsize\vspace*{-0.1ex}} \newcommand{\beas}
{\small\begin{eqnarray*}} 
\newcommand{\eeas}{\end{eqnarray*}\normalsize\vspace*{-0.1ex}}

\begin{document}


\thispagestyle{empty}
\renewcommand{\thefootnote}{\fnsymbol{footnote}}

\setcounter{page}{0}
\begin{flushright} UM-TH/97-19\\
hep-ph/9710257
\end{flushright}

\begin{center}
\vspace*{1.6cm}
{\Large\bf
The Physics of the Ultraviolet Renormalon }

\vspace{1.5cm}
{ R. Akhoury}
\vspace{0.6cm}
and
{ V.I.Zakharov}

{\it Randall Laboratory of Physics\\
University of Michigan\\ Ann Arbor, Michigan 48109-1120}\\[0.6cm] 

{\bf Abstract}\\[0.6cm]
\end{center}

We review the physics of the ultraviolet renormalon.
This mini-review is intended to be 
a sequel to the review by the same authors
at the "QCD '96" conference last year.

\vspace{1.4cm}
\vspace{0.8cm}
\newpage
{\bf THE ULTRAVIOLET RENORMALON.}

 Last-year's review \cite{az} did not cover this topic at all,
mostly because the physics of the ultraviolet renormalon
is quite different and much less transparent than that
of the infrared renormalons. Very recently, however, there has 
appeared some preliminary evidence \cite{marchesini},
obtained with
the lattice simulations,
that UV-renormalon type effects may be significant.
In another development, these effects were reconsidered within
a general dispersion relations framework \cite{grunberg}. 

{\bf Basic facts}. 

The best-known fact about UV renormalons in QCD \cite{thooft} is that
they dominate the perturbation expansions in $\alpha_s(Q^2)$ at
large orders $n$ :
\be
f_{pert}=\sum_na_n\alpha_s^n
=c_{renorm}\sum_n(-1)^nn^{\beta}n!b^n_0\alpha_s^n\label{largen}
\ee
where $f_{pert}$ is a generic perturbative series for
an observable $f$ and $c_{renorm}$ is a constant. The $n!$ behaviour
of the expansion coefficients  $a_n$ 
can be inferred from the simplest renormalon chain \cite{thooft}.
Evaluation of $\beta$ requires evaluation of loop corrections
and will be neglected in this review.
Moreover, Feynman integrals 
assocociated with the renormalon chain are dominated by very large 
$k^2_{eff}$:
\be
k^2_{eff}~\sim~e^nQ^2,\label{virt}
\ee
where $Q^2$ is assumed to be large by itself.
Because of the sign oscillations in (\ref{largen}) 
the series is Borel summable. Putting for simplicity $\beta =0$,
\be
\sum_na_n\alpha_s^n~\sim~\int_0^{\infty}{exp(-t/b_0\alpha_s)\over 1
+t}dt\label{borel}
.\ee
There exist also other ways to circumvent the divergence of
perturbative expansions associated with the UV-renormalon (\ref{largen}).
The asymptotic nature of the expansion (\ref{largen}) can be characterized by
the magnitude $\Delta$ where
\be
\Delta~=~\mid a_n\alpha_s^n\mid_{min}
\ee
and the minimization is understood with respect to $n$, with $\alpha_s$ fixed.
It is easy to see that if an expansion in 
$\alpha_s(Q^2)$ is used, then
\be
\Delta (\alpha_s(Q^2))~\sim~{\Lambda_{QCD}\over Q^2}\label{d1}
.\ee
On the other hand, if the coupling is normalized at
$\mu$ and an expansion in $\alpha_s(\mu^2)$ is considered then 
\cite{bz}:
\be
\Delta (\alpha_s(\mu^2))~\sim~{\Lambda_{QCD}^2\over Q^2}{Q^4\over \mu^4}
\label{d2}\ee
which implies that by choosing the normalization point 
$\mu^2$ high enough one can make the uncertainty of
the perturbative expansion arbitrarily small.

The last observation which we would like to mention 
among the facts concerning
UV renormalons is that the evaluation of the coefficients $a_n$ is most
straightforward if one uses an operator product expansion utilizing 
the fact that $k^2_{eff}\gg Q^2$ \cite{parisi,vz,kivel}. Moreover, operators of
dimension $d=6$ are relevant to the leading UV renormalon (\ref{largen}).
We shall give an example of
such an operator later and now notice that in this way one can
prove \cite{vz} that actually multirenormalon chains 
produce the same asymptotics
as a single chain (\ref{largen}). Thus, by $c_{renorm}$ in Eq. (\ref{largen})
one should actually understand a sum over contributions of graphs
with various numbers of renormalon chains:
\be
c_{renorm}~\rightarrow~\sum_n c_{n~renorm}.\label{collapse}
\ee
Thus a new sum is introduced and the convergence properties
of this sum are not known.
Moreover, "towers" of renormalon chains
can be important as well and the corresponding $k^2_{eff}$ can be even
much
larger than indicated by (\ref{virt}). 

{\bf Puzzles.}

The Borel summation 
brings in some puzzles as well. 
Indeed, the procedure implies that the sum over the rising branch 
of the perturbative expansion is equal to (one half of) the minimal
product $a_n\alpha^n_s(Q^2)$ which is of order $\Lambda^2_{QCD}/Q^2$.
On the other hand, applying general dispersion relations to the quantity
$f$ (whose perturbative expansion we are analyzing) one would
 conclude that the $1/Q^2$ piece is associated with resonances.
This kind of logic is behind the QCD sum rules \cite{svz} but in that
case resonances are dual to the IR sensitive part of perturbative graphs
parametrized in terms of vacuum condensates. Now, UV renormalons
bring a duality between terms 
$\Lambda^2_{QCD}/Q^2$ in the dispersive representation
and Feynman graphs at momenta of order $Q^4/\Lambda^2_{QCD}$
which looks puzzling \cite{z}.

Turning next to Eqs. (\ref{d1}), (\ref{d2}) we note that for $\mu^2\gg
Q^2$ the perturbative expansion in
$\alpha_s(\mu^2)$ is uniquely defined to an accuracy
much better than $\Lambda^2_{QCD}/Q^2$. Moreover, 
if the whole procedure is self consistent
the expansion in
$\alpha_s(\mu^2)$ appears to converge to the Borel sum of the
perturbative expansion in $\alpha_s(Q^2)$. In other words, it appears 
as if one can prove the validity of the Borel summation, which
seems to be too strong a result to emerge without extra hypotheses. 

Borel summation apparently implies a kind of generalization of the 
renormalization group to the power-like corrections \cite{parisi,cecio}.
Indeed let us consider a large but finite ultraviolet cut off
$\Lambda_{UV}$. Then we can write the bare Lagrangian as
\be
L_{bare}~=~L_4(\alpha_{s,bare})+\sum_i {c_i\over \Lambda_{UV}^2}
O_i^{(6)}\label{six}
\ee
where $L_4$ is the standard Lagrangian containing operators of
dimension $d=4$ while $O^{6}_i$ are all possible operators of
dimension $d=6$. Moreover, the standard renormalization group argument
produces a relation between $\alpha_s(Q^2), \alpha_{s}
(\Lambda_{UV}^2), 
\Lambda_{UV}/Q$. Similarly, the ultraviolet renormalon (\ref{largen})
is related to the insertions of the d=6 operators (\ref{six}).
Indeed let us estimate, by means of Eq .(\ref{virt}), the order
of perturbative expansion which is affected by the UV cut off,
\be
N~\sim~ln{\Lambda^2_{UV}\over Q^2}\label{crucial}
.\ee
Then the corresponding change in the Borel sum (\ref{borel})
is of order:
\be
\int_{(ln\Lambda^2_{UV}/Q^2)/(lnQ^2/\Lambda^2_{QCD})}^{\infty}
{exp(-t/b_0\alpha_s(Q^2))\over 1+t}dt~\sim ~ {Q^2\over \Lambda_{UV}^2}
\label{crucial1}\ee
which corresponds to the matrix elements of $d=6$ operators
over states characterized by large momentum $Q$ (unlike the
OPE used in the infrared region the matrix elements of the
operators used to evaluate the UV renormalon are calculable
perturbatively, see, e.g., \cite{vz}).

This argument leads us to believe that the "irrelevant" operators 
of $d=6$ in Eq. (\ref{six}) can be introduced only in conjunction with
the UV renormalons. On the other hand, there are no theoretical
means in fact to fix the constant in front of the coefficients 
$a_n$ due to UV renormalons. The reason was already mentioned above:
multi-renormalon chains produce the same asymptotics
as in Eq. (\ref{largen}). 
Thus, the coefficients $c_i$ could be fixed only
in terms of all $c_{n~renorm}$. However, these multi-renormalon
chains are associated with momentum $k^2_{eff}$  
even higher than (\ref{virt}) \cite{vz}. Thus Eq. (\ref{crucial}) 
does not hold generally speaking beyond one-renormalon chain
and Eq. (\ref{crucial1}) does not extend much beyond one renormalon
chain and the assumption on the Borel summability appears to be
a not-well-understood
constraint.

All these questions can well be resolved upon further analysis.
However, the time might be ripe for speculations as well.
 
{\bf Speculations.}

There are recent speculations \cite{grunberg}
which turn the puzzles discussed above into positive statements.
Namely a $1/Q^2$ non-perturbative correction is postulated to
exist in the running coupling $\alpha_s(Q^2)$ itself
and then these corrections make the question on the Borel summability
of the UV renormalon somewhat irrelevant because
there is an extra source of $1/Q^2$ terms. The argument is based on a 
dispersion
representation for the running coupling $\alpha_s(Q^2)$
and goes back to ideas expressed in the fifties \cite{redmond}.
Namely the running coupling $\alpha_s(Q^2)$ in the leading log
approximation
$$
\alpha_s(Q^2)~=~{1\over b_0ln(Q^2/\Lambda^2_{QCD})}
$$
contains a pole at $Q^2=\Lambda^2_{QCD}$ which is not present 
at any finite order of perturbation theory. If one removes this pole
from the imaginary part and invokes analyticity then one arrives at
(for a recent discussion see \cite{shirkov}) a modified expression
for the running coupling:
\be
\bar{\alpha}_s(Q^2)~=~{1\over b_0ln(Q^2/\Lambda^2_{QCD})}
+{\Lambda^2_{QCD}\over b_0(\Lambda^2_{QCD}-Q^2)}\label{q2c}
\ee
which at large $Q^2$ differs from the standard 
coupling by a $1/Q^2$ term. Although such terms are not
detectable formally via pereturbative expansions,
common wisdom tells us that the nonperturbative corrections
start with $Q^{-4}$ corrections since the lowest dimension of
a gauge invariant operator, $(G_{\mu\nu}^a)^2$ is four.
Eq. (\ref{q2c}) does introduce therefore a new kind of non-perturbative
correction at large $Q^2$ based on the idea of the duality of
the $1/Q^2$ corrections at large $Q^2$ to the region 
of small $Q^2$, $Q^2\sim\Lambda^2_{QCD}$
in the dispersion representation. 
Let us note that the reasoning for the introduction of the $1/Q^2$ correction
by itself 
is not compelling in fact. One can always impose relations on the imaginary
part in such a way as to get rid of the $1/Q^2$ term
at large $Q^2$
(see, e.g., \cite{dmw}). 

Note also that according to the logic outlined above
the $1/Q^2$ term
in $\alpha_s^2(Q^2)$ does not reduce directly to that in 
$\alpha_s(Q^2)$ and therefore the $1/Q^2$ correction cannot be 
removed by a mere
redefinition of the coupling.
Indeed
originally we have an expansion in $\alpha_s(Q^2)$
and are removing now the corresponding single pole in the perturbtaive
expression for each term. For example:
\be
{1\over ln^2Q^2/\Lambda^2_{QCD}}~\rightarrow~
{1\over ln^2Q^2/\Lambda^2_{QCD}}-{\Lambda^2_{QCD}\over
Q^2-\Lambda^2_{QCD}}\label{q2c2}
,\ee
and there is no small factor $\alpha_s(Q^2)$ in front of the
$1/Q^2$ correction
despite the fact that we started with $\alpha_s^2(Q^2)$. 
Therefore, all terms in the $\alpha_s(Q^2)$
expansion give comparable contribution to the $1/Q^2$ correction.
This collapse of the whole perturbative expansion as far as
power like corrections are concerned is similar
to collapse of contributions of all the UV-renormalon chains
mentioned above
(see Eq. (\ref{collapse})).

{\bf Physical picture}.

Although the speculations above did allow us
to settle the puzzles outlined in subsection 2,
it is an absolutely an open question whether these specualtions
are correct. The point which we emphasize here is that
if the non-perturbative $1/Q^2$ corrections
in $\alpha_s(Q^2)$ do exist, this would signify
new physics, not a simple definition or redefinition 
of the coupling coupling.
To substantiate the point let us consider \cite{paz} for a moment,
a very different problem at first sight, that is the static potential
between a heavy quark and antiquark at short distances.

At short distances the potential is 
dominated of course by a Coulomb-like piece
so that the actual problem is power-like corrections to it. 
These corrections have been considered many times (see, e.g.,
\cite{potential}) and  various approaches so far have given
similar results:
\be 
\lim_{r\rightarrow 0} V(r)~=-{C_F\alpha_s(r)\over r} + 
c_2 r^2\Lambda^3_{QCD}
\label{potent}\ee
where $C_F=4/3$ and we consider the potential in the color-singlet
channel.
Note that short distances,
that is $r\ll \Lambda^{-1}_{QCD}$ are considered.
At large distances $V(r)$ is a confining, linear in $r$ potential.

Although each time \cite{potential} 
there is a particular model behind Eq. (\ref{potent})
the physics can be explained in a very simple way
\cite{paz}.
Namely, consider 
an Abelian case and represent the potential as an integral over 
electric fields of two charged particles:
\be
V(r)~=~{1\over 4\pi}
\int d^3r'{\bf E}_1({\bf r}')\cdot{\bf E}_2({\bf r+r'}). \ee
In particular, the Coulomb potential can be obtained of course by 
integrating
over the fields ${\bf E}_{1,2}$ of two point charges. On the other hand, if
the electric fields are modified at large distances, then there arises a correction
to the Coulomb energy at small distances as well.

Consider as an example two charges of opposite signs in a cavity 
of size $R$, $R\gg r$.
 Then the electric field of the charges, which is that of a dipole at large
 distances in empty space, changes at $r'\sim R$. The corresponding change in
$V(r)$ is of order
\be
\delta V(r)~\sim~\alpha r^2\int_R^{\infty}
{d^3r'\over(r')^6}~\sim~{\alpha r^2\over R^3}\label{cavity} \ee
which is in agreement with the correction to the static potential
above. Indeed, in QCD one just assumes that the 
perturbaive fields are changed 
at distances $R_{cr}\sim\Lambda_{QCD}^{-1}$ where the running coupling
$\alpha_s(R_{cr})$ becomes of order unit.

Let us introduce now an alternative  
model according to which the (color) electrostatic field of quarks
is a correct zeroth-order aproximation only as far as it exceeds some critical
value of order $\Lambda_{QCD}^2$: \be
({\bf E}^a)_{cr}^2~\sim~\Lambda_{QCD}^4 \label{weak} \ee
while weaker fields do not penetrate the vacumm because of its specific,
confining
properties. From this condition we get an estimate of distances $R_{cr}$ where
 the electrostatic field of quarks is strongly modified: \be
{\alpha_s r^2\over R_{\rm cr}^6}~\le~\Lambda_{QCD}^4 \label{ratio}\ee
where for simplicity we have neglected the effect of the running of
$\alpha_s (r')$.
The change in the potential is then of order: \be
\lim_{r\to 0}\delta V~\sim~{\alpha_sr^2\over R_{\rm cr}^3}~\sim~\alpha^{1/2}r
\Lambda_{QCD}^2 \label{nnew},
\ee
i.e., we get a linear in $r$ leading correction to the potential at short
distances.

Now, we can turn back to the $1/Q^2$ correction to $\alpha_s(Q^2)$.
The point is that such a correction would also bring  
a linear correction to the potential at short distances
and it is natural to speculate that the mechanism behind this
$1/Q^2$ correction is the same as discussed above in the case of the
potential.

Although condition (\ref{weak}) might look very natural
it is worth emphasizing that to realize such a condition 
we need small size non-perturbative fluctuations in the
physical QCD vacuum. Indeed, according to (\ref{ratio})
$R_{cr}\rightarrow 0$ if $r\rightarrow 0$.
If one tries to speculate, what kind of fluctuations 
these could be, it is natural to turn to the dual superconductor
picture of confinement \cite{mandelstam} (for a recent review see
\cite{baker}). Magnetic monopoles are a crucial field configuration in this
case. The magnetic monopoles of QCD were introduced \cite{thooft1}
in the Abelian projection of QCD where they appear as singular objects.
Although this could be an artifact of the gauge fixing \cite{thooft1},
convincing evidence for existence of monopoles as physical objects was
obtained in just this gauge (see \cite{polikarpov} and references therein).
If the physical size of monopoles is indeed vanishing the linear 
potential at large distances, could well continue to $r\rightarrow 0$.
This can be inferred, for example, from the results of ref.(\cite{japs}).     
One may even say that measurements on the lattice of the potential at short
distances could give meaning to the notion of a monopole size.

{\bf Phenomenology.}

The phenomenology of the UV renormalons is in its infancy.
There exists only a single experimental indication \cite{marchesini}
obtained on a lattice
that $\alpha_s(Q^2)$ does receive a $1/Q^2$ correction at large $Q^2$:
\be
\alpha_s(Q^2)~=~{1\over b_0lnQ^2/\Lambda^2_{QCD}}+c_{lat}{\Lambda^2_{QCD}\over
Q^2}
\label{lattice}\ee
with $c_{lat} >0$. Note that this positiveness of $c_{lat}$
rules out the most naive identification with the $1/Q^2$ correction
in Eq. (\ref{q2c}). Beyond this naive level, however, even this 
observation should be taken with caution.
Indeed as is discussed above (see Eq. (\ref{q2c2}))
the $1/Q^2$ corrections due to all orders in $\alpha_s(Q^2)$ 
are of the same order
and Eq. (\ref{lattice}) should be 
taken rather as a neumonicfor the
results of the mesaurements 
of $<\alpha_s(G_{\mu\nu}^a)^2>$ than a real determination of $\alpha_s(Q^2)$
with inclusion of power like corrections.
It would be of great importance of course to further check that
nonperturbative corrections at large $Q^2$ start with $1/Q^2$.

As for direct measurements of the $V(r)$ on the lattice (see \cite{bali}
and references therein) they do not indicate any change of linear
in $r$ behaviour at large $r$ to the $r^2$ behaviour at short distances
predicted by the "standard" QCD (see Eq. (\ref{potent})) and in this sense
one might say that expectations of a linear correction at short distances
are confirmed (see Eq. (\ref{nnew})). However existing measurements 
do not target power corrections to the Coulomb potential at short distances
specifically and no statement on these corrections has been made.

Direct evaluation of ultraviolet renormalon contribution
to various observables was tried in a few papers, see, e.g. \cite{sf}.
The result is scheme dependent and usually does not represent
a leading power correction. It is more in the spirit of the
present review to consider rather 
 UV-renormalon inspired
phenomenology. Namely one starts with UV renormalon contributions
to various quantities and assumes, explicitly or tacitly,
that they are enhanced in some way to a phenomenologically singnificant
level. The approach is similar to that one widely practiced for
IR renormalons (for a review see \cite{az}). And although such
a phenomenology can be truly successful only if
non-perturbtive $1/Q^2$ corrections exist and are relatively large,
it can be developed without discussing the mechanism of the enhancement.
There are some encouraging qualitative results obtained.
First, the UV renormalons are universal in the sense 
that their contributions are determined
in terms of anomalous dimensions of a few 
operators of dimension six \cite{vz,kivel}.
These operators can be split furthermore into operators which
can be added directly to the bare QCD Lagrangian (see Eq. (\ref{six}))
and operators which are process dependent. The latter set
depends, for example, on whether one considers correlator of
vector or pseudoscalar currents. It turns out that
the process-independent operators dominate.   
It is not trivial since these operators appear first on the 
two-renormalon-chain
level while process-dependent operators can be associated already with
a single chain.
Moreover, the quark operators are then the same that are postulated
within the Nambu-Jona-Lasinio model
indicating for the first time a possible connection between
this phenomenological model and fundamental QCD \cite{vz,yamawaki,peris}. 

Moreover, one may try to explain \cite{yamawaki} by inclusion of
$1/Q^2$ terms
the irregularities \cite{novikov} in the QCD sum rules in various channels.
Essentially, in the same spirit as described in subsections 2,3
one speculates now that UV renormalons are dual to the pion
(for other possible explanantions 
of the failure of the standard QCD sum rules in the
pion channel see \cite{novikov,shuryak}).
On the other hand, there can be no such duality for the $\rho$
\cite{narison}. For this to be true, the leading quark operator 
of dimension six should be
\be
O^{(6)}_{lead}~=~(\bar{q}{\bf \tau}^i\gamma_5 q)
(\bar{q}{\bf\tau}^i\gamma_5 q)+(\bar{q}{\bf \tau}^i q)(\bar{q}{\tau}^iq)
\ee
where $\tau^i$ are the Pauli matrices in the flavour space of $u$ and $d$
quarks. This difference between the $\rho$- and $\pi$- channels is
a nontrivial check of the whole scheme.
It is amusing therefore that a direct 
calculation of two renormalon chains \cite{peris}
does indicate this difference:
\be
(UV~renormalon)_{PS}~=~18(UV~renormalon)_V 
\ee
giving hopes for the relevance of renormalons. 

{\bf Conclusions.}

One might say that there is a recent trend to 
view the UV-renormalon physics
in a way similar to the IR physics.
Namely one assumes that nonperturbative fluctuations 
produce $1/Q^2$ corrections which enhance the 
UV renormalon contributions.
This analogy between UV and IR physics is far from trivial
since it assumes the existence of small-size nonperturbative
fluctuations in QCD. This strategy may well be wrong.
On the other hand, if it is confirmed  by 
measurements it would provide 
new insight into the physics of confinement.

The authors are grateful to P. van Baal, G. Marchesini, 
S. Narison and A.I. Vainshtein
for useful discussions.

\end{document}